# Large Photocathode Photodetectors Using Photon Amplification and Phase-Space Compression

*Alex Carrio[1], Joseph Dowling[1], Kevin Greener[1], Sean McGuiness[1], Victor Podrasky[1], John Sullivan[1], David R Winn[1]\*,
Burak Bilki[2], Yasar Onel[2]*

1. Department of Physics, Fairfield University, Fairfield, CT 06824 USA
2. Department of Astronomy & Physics, University of Iowa, Iowa City, IA, USA

\*Corresponding – winn@fairfield.edu

*Abstract:*
We describe a simple technique to both amplify incident photons and compress their angular x area phase space. These Optical Compressor Amplifier Tubes (OCA Tube) use techniques analogous to image intensifiers, using vacuum photocathodes to detect photons as converted to photoelectrons, amplify the photons via photoelectron bombardment of fast scintillators, and compress the optical phase space onto optical fibers, so that small, high gain photodetectors, like miniature PMT or SiPM, can be used to detect photons from large areas, at comparatively low cost. The properties of and benefits of OCA tubes are described.

*Introduction:*
Photomultiplier tubes (PMT) and SiPM are ubiquitous in particle detection apparatus and experiments. Their virtues are the ability to generate detectable signals from one photon, at high speed. In general, PMT gain-bandwidth is still unmatched. Experiments planned for high energy physics, particle astrophysics, intensity frontier and intermediate energy and nuclear physics anticipate the need for:

a) Large areas of photocathode for non-accelerator experiments such as for nucleon decay, neutrino oscillations, or large underwater or ice detectors for astrophysical phenomena.
b) Cosmic gamma ray telescopes and cosmic ray detectors, both terrestrial, and in satellites, which need to operate at low power, low mass, remotely, and over long times.
c) Operation of photodetectors for calorimeters in high magnetic fields;
d) Large numbers of photodetector channels such as for scintillator calorimeters[1,2] including neutrino oscillations, or for ring-imaging Cherenkov counters (RICH)[3].
e) Radiation Hardness is an issue in many experiments.

Photodetectors for these experiments have requirements which may not be entirely available with existing photodetectors. These requirements include: extended Dynamic Range and Linearity, Cost and Channels (cost of large numbers of channels are often inhibiting factors in the design of large detectors); Size, Power, Cables and Cable-Driving; Magnetic Field Operation; Gain Stability: The secondary emission gain mechanism is reasonably stable, yet it is unusual for photomultiplier (PMT) systems to have gain stability without periodic calibration better than 1-2%, and largely have low outputs in field larger than 1 T. SiPM work in high magnetic fields, but radiation damage more easily than PMT and have temperature and voltage stability requirements.

In order to help reach the goals outlined above for future generations of particle detectors and for novel photodetector applications, we developed a photodetection technique which amplifies and "compresses" phase space of the incident light on the detector, with sufficient compression on the output of the light amplifier to match the input phase-space of optical fibers. Optical fiber(s) then transport the amplified/compressed light to pixel-sized detectors (such as avalanche photodiodes (APD), SiPM, multichannel or miniature PMT) either on the OCA tube, or even in remote locations. Configurations of this Optical Compressor-Amplifier (OCA) might be used to:
(a) obtain useful photon gain from very large area photocathodes at low cost;
(b) operate densely packed highly compact vacuum photodetectors in confined spaces with minimal cable and service corss-sections, without the need for cable driving preamps;
(c) operate vacuum photocathodes in configurations where the photoelectrons can be used for gain even in the presence of large axial magnetic fields.

*The Essential Technique for an Optical Compressor Amplifier:* The basic technology for light amplification and compression is old: photoelectrons from a standard vacuum photocathode are accelerated through a large voltage and are then reconverted back to photons on the anode consisting of an aluminized fast phosphor, as in an image intensifier. The photoelectron image can also first be demagnified, compressing the areal density. We propose to use this old technique in some new geometries, and with very fast near UV/blue phosphors, with the phosphor light coupled to wavelength shifting fibers, which make this technique useful for new types of optical detectors. Essentially, we propose an image intensifier, where we throw the image away and just collect the amplified light, but in the especially convenient form of light compressed and coupled into optical fibers. Small photodetectors, optionally in remote locations are then used to generate the electrical signal corresponding to the incident photons. The novel methods we propose to introduce are:

*(a) Planar:* Photoelectrons impinge on a proximity focussed 1:1 cathode-anode area, short wavelength phosphor (blue or UV) plate, and the resulting photons are then collected from the phosphor anode plate via a loop or loops of wavelength shifting (WLS) plastic fibers inserted into a groove in the bottom glass of the tube, or in a polymer (plastic) matching plate. This WLS readout is similar to detection schemes in CMS[4] and ATLAS calorimeters at LHC. The WLS fiber needs to cover no more than a few percent of the area of the scintillator for efficient coupling and uniform collection. This arrangement could function even in a large axial magnetic field as the incident angle of the photoelectron is not critical. Figure 1 shows a schematic of this photoamplifier, and the WLS fiber output.

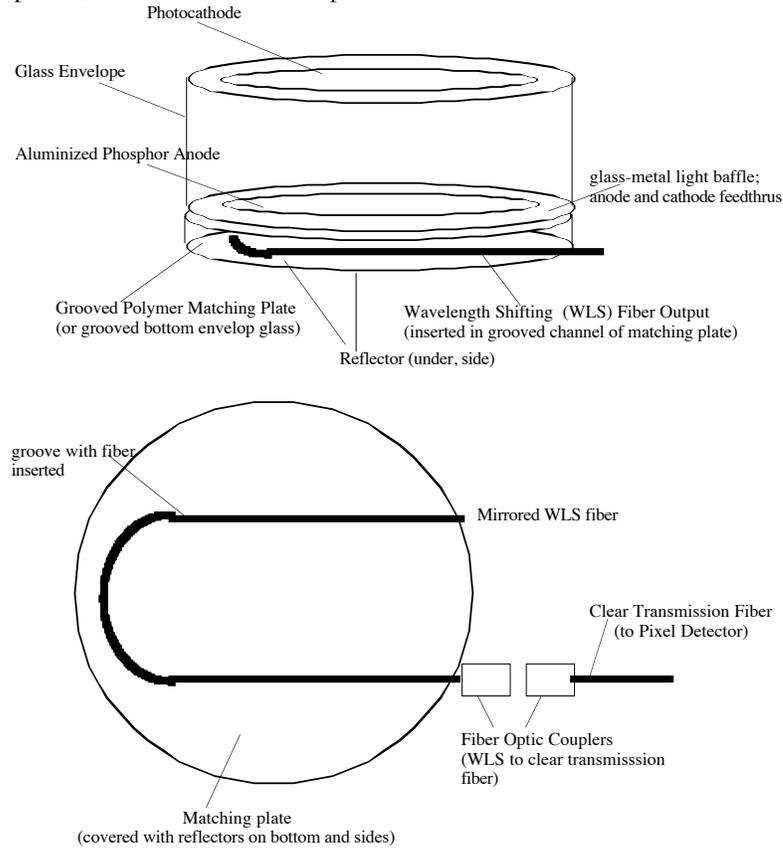

**Fig. 1:** *Top:* Schematic of compact proximity-focussed optical compressor amplifier (OCA) using a wavelength shifting (WLS) optical fiber output, for possible use in high axial magnetic fields. Anode-Photocathode spacing: 0.25-0.1 mm/KV. Not shown: details of photocathode and phosphor anode electrical connections (similar to generation-1 image intensifiers). *Bottom:* Plan-view of possible matching plate, groove, and WLS fiber output. WLS fiber could be looped in many configurations, including a tightly would spiral for efficient collection

*(b) Coaxial Cylinder:* A variation is to use a phosphor-fiber hollow capillary tube as the anode, and insert a WLS fiber in the hollow interior. Figure 2 shows a schematic of this configuration. The vacuum volume forms a "donut" shape, with the interior of the phosphor anode open to the air. This is essentially the configuration tested in Figs 4. The photoelectons travel from the outer photocathode cylinder to the inner phosphor cylinder. This configuration can scale to arbitrarily long lengths depending on the attenuation of the WLS fiber, which would be coupled to clear fibers at both ends. (In principle, time and charge division could be used to localize the centroid of the photoelectrons.) Using existing green WLS fibers, 2-3 m long tubes for large underwater experiments might be fabricated. The coaxial cylinder configuration has the advantage of being capable of pressure standoff (like a submarine geometry), but uses far less volume for a given area than a spherical configuration (like the large Hamamatsu PMT used in SuperK) where most of the vacuum volume is unused, in effect. The photoelectrons are radially focused, and the trajectory angle at the phosphor is not critical as it is with dynodes. Remarkably, the photoelectron trajectories are nearly isochronous - intrinisic time jitter is minimized to sub-ns values, and the time error is dominated by the phosphor decay properties, which can be made small with at least 2 well-known phosphors.

By "compressing" the phase space of detected photons in this manner, it is possible to take advantage of lower gain, smaller and inexpensive photodetectors/photomultipliers, or pixel detectors, for reading out a large number of photon amplifiers. As long as the transmitted photons per photon is larger than 1, this technique for amplified light detection has been shown to work[5] as a compressor, and also in several other contexts also similar to image intensifiers,[6,7,8,9]. We propose to extend this proven technique by producing the concentrated and amplified photons directly in a fiberoptic-coupling structure.

The angle and position of the accelerated photoelectrons impinging on the phosphor plate or cylinder is not critical compared with the trajectories of photoelectrons in most secondary emission amplification devices, and so operation in a large magnetic field oriented along the photoelectron direction is acceptable (see discussion below).
.

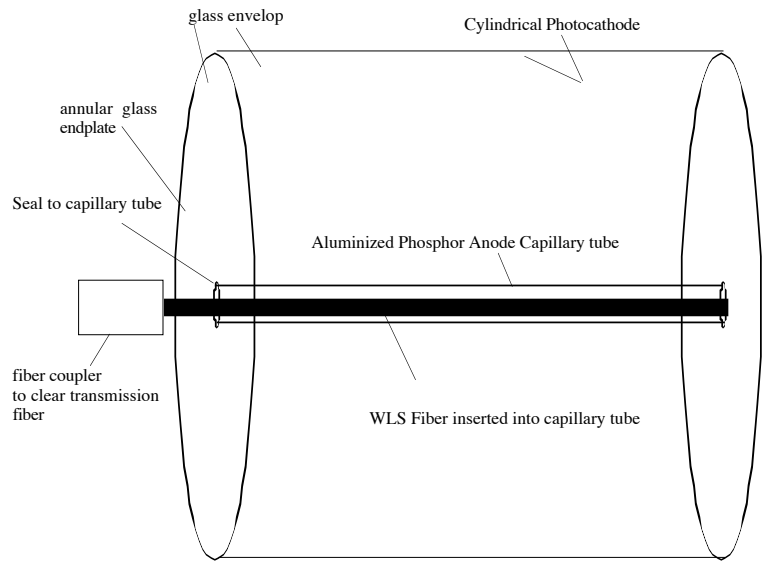

**Fig. 2:** Cylindrical optical compressor-amplifier (OCA) - A cylindrical aluminized hollow annular aluminized phosphor anode configuration, for readout of a photocathode of radius R using a coaxial WLS fiber inserted in the interior of the anode (radius r). The vacuum space is a stretched torus; the inner, small glass cylinder is open to the air on its inside (ID), and vacuum on its outer edge, or OD. In long aspect configurations (for example, like a fluorescent lamp tube) it may be essential to read out both ends of the WLS fibers.

*Phase Space Compression:* The compression of the photon phase incident on the OCA tubes comes about by 2 methods: the photoelectron trajectories, and by the collection using wavelength shifting. There are 2 extreme cases, and hybrids using combinations both of them can be obtained.

*Radial-Focussed Case (Fig 2):*
This case is the convenient limiting case for obtaining very large photocathode tubes, in the absence of appreciable magnetic fields. In the absence of a magnetic field, photoelectrons emitted from a photocathode of radius R will cross an anode of smaller radius r roughly according to

$$r \sim R(E_{pe}/V)^{1/2} \qquad (1)$$

(up to constants close to unity), where $E_{pe}$ is emission the energy of the photoelectron (~0-2 eV) and V is the voltage in Volts of the anode-cathode difference.(This relationship corresponds roughly to angular momentum conservation). A 40 KV tube could have an photon areal phase space compression exceeding $5 \times 10^4$. A 5 cm diameter cylindrical photocathode could deposit 40 KV photoelectrons on a ~200$\mu$ diameter phosphor anode cylinder, or a 12" cylindrical cathode could deposit photoelectrons on a ~2 mm diameter phosphor cylinder (the deposited photoelectrons then lose memory of the electron trajectory, which is irrelevant in the phosphor). Further compression can then be obtained with the coaxial nested WLS fiber.

*Proximity Focussed (Planar Cathode) Case:* This case is the limiting case for obtaining gain in very high magnetic fields. In this case (Fig. 1), the photoelectron phase space has the same area as the incident light, but the photoelectrons have a fractional angular spread which is somewhat more limited than the incident light, due to the proximity focus. . The phase space compression factor K can be approximated by the product of the ratios of areas and angles of the photocathode and the fiber:

$$K = (D_{pc}/D_f)^2 \times (\sin\theta/N.A.) \quad (2)$$

where ($D_{pc}$ and $D_f$ are the diameters of the photocathode and the output fiber and $\sin\theta$ is the maximum angle that the incident light hits the photocathode and N.A. is the numerical aperture of the fiber. This can easily exceed 10,000 in a practical case. The limitation to phase space compression of the light incident on the face of the tube is mainly the reflectivity and self absorption of the phosphor plate assembly (and also the transmission of the the fiber). The light emerging from the phosphor must be trapped in the plate assembly until it is shifted and captured by the fiber. As the reflectivity of the top aluminization decreases and self-absorption increases, the WLS fiber must be made longer to cover more of the area of the phosphor. (Alternatively, a number of smaller WLS fibers could be arrayed on the bottom of the tube, and coupled to a larger clear output fiber - for example seven ~300 $\mu$m WLS fibers could be coupled to a 1 mm clear output fiber.) Self-absorption thicknesses for typical phosphors (about 0.02 g/cm$^2$ [10]) is usually much larger (by a factor of 10-20) than the minimum necessary phosphor thickness, given by the electron range (see discussion below).

*Photon Gain:*
*Phosphor Anodes:* The photon gain per photoelectron depends on the cathode-anode voltage difference and the phosphor. A convenient way to parametrize this is g, the number of photons produced per KeV deposited in the phosphor by the accelerated photoelectron. For example, a NaI anode could in principle produce ~40 photons/KeV but would radiation damage. Fortunately, high efficiency phosphors are available as developed for CRT, image intensifiers, streak cameras and fluorescent lighting, and some of them are sufficiently fast decays to be interesting in detector applications. ZnO(Ga)[11],[12],[13] would give the best time resolution (0.4-0.75 ns decay, 40-60photons per KeV – up to 1.5 times NaI, 390 nm peak wavelength)[14],[15],[16]. CdS:In has 525 nm peak emission, <1 ns decay to 10%, 0.5 light output of NaI and no afterglow. Newer nanocrystalline phosphors less than 50 nm in diameter, typically ~100 atoms, with an exciton wavelength larger than the nanoparticle, decay in less than 1 ns, with energy efficiencies exceeding 50%. A typical phosphor film thickness does not need to exceed ~5 microns to efficiently stop the electrons up to ~40 KV. The penetration depth T ($\mu$m) of the electrons into typical phosphor films is given approximately by:

$$T = 1.1 \times 10^{-6} V_b^{1.4} \ \mu m \quad (3)$$

where $V_b$ is the incident electron (photocathode to anode) voltage. At 50 KeV, a typical electron range is ~ 0.001 g/cm$^2$ [17].

The phosphor films are overcoated with an aluminum film to a thickness of about 50-100 nm (an optical density of about 11), but remain basically transparent to the energy of multi-KV electrons over a large range of angles. The optical density is large enough to prevent optical feedback to the photocathode, and the aluminum layer is sufficiently conductive also to conduct the photocurrent and maintain the cathode-anode electric field. On average, typically ~400-700 eV of a multi KeV photoelectrons are lost to the aluminizing in an image intensifier, integrated over the range of angles of the photoelectron incident. The reflectivity is such that the light emerging from the outer side of the phosphor is increased by a factor of 1.8-1.9 in a typical phosphor screen, indicating an effective reflectivity and self-transmission up to 90%, including the mismatches of the index of the phosphor with the glass of the envelope.

The anodes can be formed with a thin phosphor film by using: (i) the standard settling binder technique[18] (using fine phosphor powders dispersed in a silicate solution binder, which adhere to the item to be coated, which is then fired at moderate temperatures), (ii) by PVD (e-beam evaporation preferred), (iii) by sputtering, or (iv) by CVD.

*HV:* The HV is considerably higher than those used in ordinary PMT, but are typical for image intensifiers, used in night vision, and powered by batteries via a Cockroft-Walton-like voltage multiplier. The design rule for vacuum insulators is 100 KV/cm in manufactured products (precision CRT, travelling wave tubes, compact klystrons, crossatrons). Typical large-screen TV electron guns operate at 40 KV over less than 1 cm in the extraction electrode. The HV power small because the current is only the photocurrent.

*Total Photon Gain:* The overall gain G of photons captured in an output fiber per photons incident on the photocathode is given approximately by:

$$G = Q\, \varepsilon_k\, V\, g\, \varepsilon_c\, \varepsilon_f \qquad (4)$$

where Q is the quantum efficiency of the photocathode, $\varepsilon_k$ is the average collection efficiency of the photoelectrons on the anode, V is the anode-cathode voltage, corrected for loss in the aluminization, g is the phosphor light emission in photons per unit electron energy, $\varepsilon_c$ is the capture efficiency of the produced light (dominated by the aluminum mirror of the anode coating, the refractive index mismatches, and self absorption of the phosphor), and $\varepsilon_f$ is the acceptance of the fiber given by the numerical aperture. Using guesstimates of Q=20%, $\varepsilon_k$=90%, 40-50 KV, g=40-60 photons/KeV, $\varepsilon_c$=50% and $\varepsilon_f$ =4%, respectively, we obtain ~5-8 photons captured on a fiber per incident photon. This is a conservative estimate using the high speed ZnO(Ga) phosphor and a fluoride-quartz cladded quartz core fiber. Using activated aluminate phosphors (P41, P47 - like phosphors), and high NA fibers (0.44), more than an order of magnitude higher photon per photon gain may be possible (i.e. ~20 photons/photon) compressed to the phase space of an optical fiber. Assuming a high quality fiber coupling, these photons could be transported long distances to a pixel detector.

*Temporal Response:* The temporal properties are dominated by the scintillation decay of the phosphor and the WLS, as the intrinsic time-spread of the photoelectrons caused by the electron optics could be made insignificantly small in the 2 geometries discussed. A 1 ns phosphor and WLS fiber would add about 2-3 ns to the width of the readout detector pulse, and using the 10% rule of thumb, less than 1 ns to the time resolution.

*Electron Optics, Size Scaling, and High Magnetic Fields:*

For a radial-focused OCA tube, as in Fig.2, the diameter is limited by the total voltage standoff, say, 80-100 KV. Using WLS fiber bundles (cylinder of fibers covering a rod) very large diameter cylindrical photocathode OCA tubes could be obtained. A 5 mm diameter WLS "rod" could read-out a 50 cm diameter, 50 KV OCA tube. The length of the OCA tubes configured as radially focused cylinders is limited by the light transmission in the WLS fiber. To obtain uniformity of readout or spatial resolution, both ends of the WLS fiber should be readout. Using ~ 1mm WLS fibers, 2-3 m long tubes could be readout for large-scale particle astroparticle physics experiments.

For a proximity-focused tube (Fig 1), with care, the design of vacuum HV standoff can achieve 50-100 KV/cm. Thus a 30 cm diameter, 40 KV, proximity focused light amplifier with a waveshifting fiber fiber-optic output could be ~1-2 cm thick (Fig. 1). In this case of the proximity-focused phosphor plate operated in high axial magnetic fields, we note again that this type of tube is simply a Gen I image intensifier, where the image is not used. Some of the larger Gen I tubes were immersed a 1-2 T axial magnetic field in order to preserve the photoelectron-image, using the magnetic field to guide the photoelectrons to the phosphor screen[19], confining the spiraling electrons to small helixes whose axes connect the photocathode to the phosphor anode in straight lines. This existing practice in many image intensifiers serves as a practical demonstration that photon gain from a photocathode to a phosphor can occur in a high axial magnetic field. A fraction of photoelectrons will always be emitted with shallow enough angles to the field direction so that the electron spirals will be small, the electrons will impinge on the phosphor anode at modest angles to the normal, penetrate the aluminization, and gain will always occur. (The differential yield of photoelectrons is proportional to $\cos^2\theta$, where $\theta$ is the angle to the normal of the cathode plane.) This is similar to but better than the situation with high magnetic field PMT using mesh dynodes or MCP where the initial electron trajectory should be be perpendicular to the plane of the gain stage. In these cases, PMT may operate up to 2 T axial fields, where they are then limited by the difficulty of the low energy secondary electron optics from stage to stage or along the microchannel impinging at grazing angles on the secondary emitter surfaces. This is because secondaries are emitted over a wide range of angles and spread laterally, becoming trapped in the field, and the voltage traversed between gain collisions is small. In case of the phosphor anode, (a) the electron accelerates to energies that can be an orders of magnitude larger than in PMT (~10 KeV compared to ~100 eV), and (b) once the photoelectron enters the phosphor, the collisionally damped dissipation of the photoelectron energy in the phosphor is only weakly affected by the imposition of a magnetic field, unlike the conditions of secondary emission, which depends strongly on the incident angles. Such a photoamplifier minimizes the active components in a very inaccessible location of the experiment, and is likely to be significantly radiation hard with proper material selection.

*Fiber Coupling:* An extensive review of fibers is given by H.Leutz[20], CERN, as shown in Fig. 3 below.

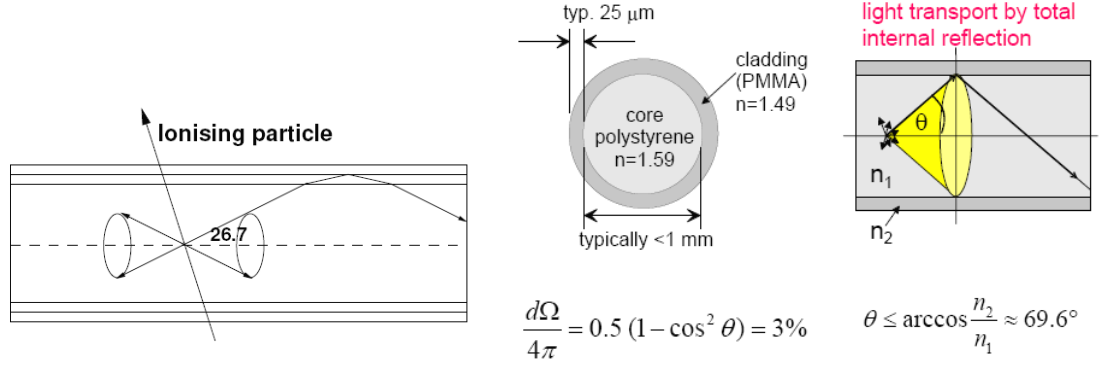

**Fig. 3:** Cartoon of light trapping in a typical plastic green WLS fiber (Leutz ref. 20).

The fraction of light piped to each end of a fiber is shown in the cartoon of Fig. 3. The fraction piped is given by the cladding $n_2$ and the core $n_1$ by:
$$f = 1/2\,(1 - n_2/n_1). \quad (5)$$
The critical angle is given by:
$$\sin \alpha = n_2/n_1, \quad (6)$$
the number of reflections in terms of the diameter d and length of fiber L by:
$$N = \cot \alpha \, L/d \quad (7)$$
and the numerical sperature NA is given by:
$$NA = (n_1^2 - n_2^2)^{1/2} = \sin \theta \quad (8)$$
For small diameter fibers, the imperfections in the walls dominate. The reflection loss length $L_R$ over which the injected intensity is reduced by 1/e from reflection imperfection losses $0<q<1$ (i.e. q is the effective the reflectivity back into the trapped region) is given by:
$$L_R \sim 1.5/(1-q)\,(n_1/NA)\,d \quad (9)$$
For many scintillators and claddings this will dominate over scitnillation light self-absorption for small diameter fibers. In general, L, the 1/e total attenuation length is given by:
$$L^{-1} = L_A^{-1} + L_R^{-1} + L_s^{-1} \quad (10)$$
where $L_s$, $L_A$ are the scattering and absorption lengths.

    Blue or green WLS (wavelength shifting) fiber loops would be coupled to the outside of the phosphor anode deposited on an optically flat glass plate or inserted into the capillary. The plate would either be thick enough to be grooved for the WLS fiber, or be bonded to a polymer matching plate with the grooves to accept the WLS fiber. A total internal reflective boundary together with a reflective covering would terminate the back face. The efficiency of the coupling will depend on the total area of WLS fiber(s), and the reflectivity. For example, WLS fibers could be wound into a spiral which would intercept nearly all of the phosphor light with no bounces. For a 1" diameter phosphor anode, ~1 m of 0.5 mm diameter WLS fiber would be required for 100% coverage. For scintillating plastic plates in calorimeters, only a few percent of the area of the plate needs to be covered with WLS in grooves for efficient capture of most of the primary light, as shown by the hadron calorimeters designed for CMS/ATLAS at CERN. This trapped light can then be efficiently coupled to clear fibers. Multiclad clear plastic fiber which can transmit 500 nm light over 50 m with 50% transmission are now commercial[21].

### *A Prototype Large Area Optical Compressor Amplifier:*

A prototype large area Optical Compressor Amplifier (OCA) device has been constructed, which collects photons over a large area, and produces more photons in a smaller phase-space[22]. It is very similar to the configuration of Figure 3, but uses a WLS rod rather than a fiber. This vacuum "image intensifier"-like device was prepared in cylindrical coaxial geometry by fabricating a small hollow (1cm diameter x 40 cm) glass tube coated on the outside with an aluminized fast phosphor powder (cylindrical "screen"). This phosphor tube is mounted coaxially inside a much larger glass tube (10 cm dia. x 40 cm long), sealed between two glass end annuli, forming a large "toroidal" vacuum region between the nested cylinders. Photoelectrons emitted from a cylindrical photocathode formed on the inside of the outer glass tube were radially focussed and accelerated to ~20 KV onto the inner phosphor cylinder, producing photon gain (all photoelectrons impinge on the "screen"). The resulting photons emitted by the phosphor are reflected into the center of the hollow phosphor tube, which is open to the air. These secondary photons are trapped in a small waveshifter bar inserted inside the small hollow phosphor tube, and piped out to a small standard phototube(s) to generate a signal. A schematic and photograph of this cylindrical cathode device is shown in figure 4a,b.

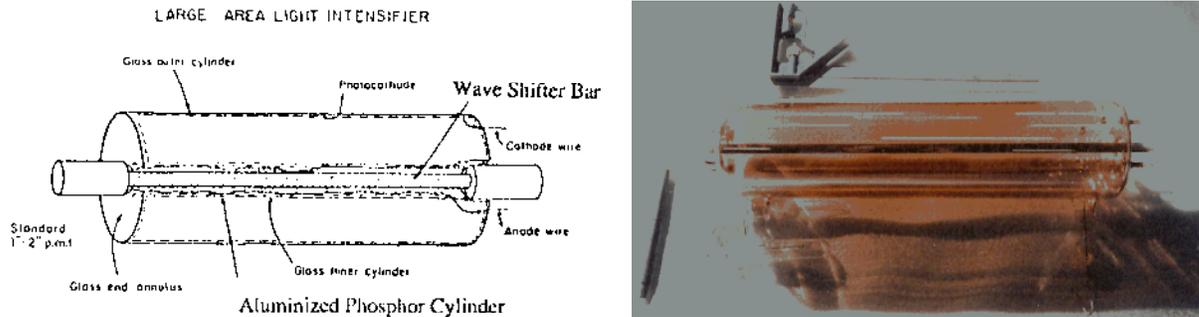

*Fig. 4a Left:* Schematic of the large area OCA (optical compressor amplifier) 40 cm x 10 cm diameter.
*Fig 4b Right:* Photograph of the prototype large area OCA prototype 40 cm x 10 cm diameter made by one of the authors. The S-11 ($Cs_3Sb$) cathode was fabricated in-situ using 6 Sb-coated W wires

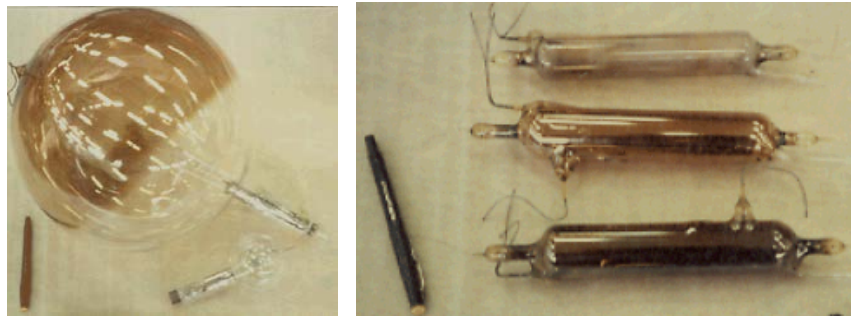

*Fig 5a,b:* 30 cm spherical photocathode diode, and b: 3 cylindrical diodes with an anode wire and highly purified 90%Ar10%Methane for gas gain. Note the 3 different thicknesses of cesium antimony photocathode. These tubes are examples very high photoelectron areal phase-space compression, the former being close to 3-D compression, and the latter 2-D compression like the OCA tube above.

Phosphor Tube and Cathode: The inner phosphor tube was made using the standard gravity settling binder technique[23]. The tube was rotated at about 1 rpm in a specially constructed Al tank where phosphor settles on the glass, which is then carefully dried and heated. A phosphor was a blue-UV emission similar to P41 (a yttrium aluminate-type phosphor) a 10 ns decay and 1/2 the light output. The phosphor tube was sent out to a company specializing in highly uniform aluminizing. The phosphor tube was then mounted inside the large pyrex envelope with 2 pyrex annuli at the ends to form a stretched evacuated torus, and the cathode formed by using 6 Sb-overplated, Au-plated W-wires stretched along and arranged around (60° intervals) the inner phosphor tube.

Cathodes and cathode materials: The basic requisite techniques for photocathode fabrication can be summarized as: a) exceptionally clean conditions (stainless steel and glass wherever possible, extensively baked out); b) a good, clean high capacity demountable vacuum system; c) a controllable oven capable of 400°C; and d) good glass-blowing technique for initial fabrication, connection to and final pinch-off of the device from the vacuum rig after fabrication, and e) long experience. For an excellent review of photocathode technology, see reference [24]. The cathode materials for Antimony-Alkali cathodes were purchased commercially. The purified antimony was evaporated and distilled onto the glass (or other) envelope by conductive heating of cage of six Sb coated wires, before activation with the alkali metal(s). A very slight modulation of the thickness resulted. The Sb source plating thickness determines the area of glass that can be coated with an Sb layer, typically 100 Angstroms thick for a semi-transparent cathode (typically 3-4 $\mu g/cm^2$ of Sb). Alkali channels (alkali salts, such as cesium chromate, with zirconia powder as a reducing agent) are obtained commercially from companies that manufacture getters (principally SAES). These "channels" are small (size of a pill) metal containers which when self-heated by an electric current produce alkali vapors.

Cathode Deposition: The Sb is evaporated at an oven temperature of 160° C at $10^{-7}$ Torr by running a current through the Sb coated wire or wire-mounted bead. The evaporated Sb vapor travels in a straight line and distills on the glass. It forms a grey metallic film which is heavy enough for a photocathode when it is 30% transparent to visible light. Antimony source geometry and shadow masks are used to ensure uniformity and proper cathode shape. The Cs (and other alkali) vapor is then reduced by a current in the channel, which quickly fills the tube, forming the stochiometric compound of the photocathode. An excess is baked off until the quantum efficiency reaches its peak value from the peak photocurrent and to test the devices, using a lamp and a picoammeter while controlling the oven and the current in the Alkali channel. Excess Cs can be baked off, or more added, and so the optimum can be found by trial and error.

Results: At 20 KV with an "intermediate" experimental phosphor (1/2 gain of P41, 10 ns decay) borrowed from Westinghouse, and using a BBQ plastic waveshifter rod coupled to 2 pmt's, we measured a gain of 5 photoelectrons on the small readout phototubes per photoelectron produced in the intensifier tube, about a factor of 3 less than calculations in a simple Monte Carlo indicated. This would indicate that about 30-35 photons/photoelectron were captured in the BBQ rod and piped to the small phototubes. The pulse decay time on the small tubes was consistent with the phosphor decay time when using large pulses of light from an Americium+Pilot-U 1.4ns scintillator source. The gain was linear with the voltage between 15-25 KV, and highly linear with incident light. We are thus confident that these results on a large cylindrical Light Amplifier support the principle of the device with fiber optic outputs we propose below.

*A Prototype Planar Assembly Optical Phase Space Compressor/Photon Amplifier:*
A prototype Planar Light Amplifier (Gen-0 image intensifier with near UV/blue output) to a matching Phase Compressor WLS fiber was designed and fabricated. The basic point design is shown in Fig 6 below.

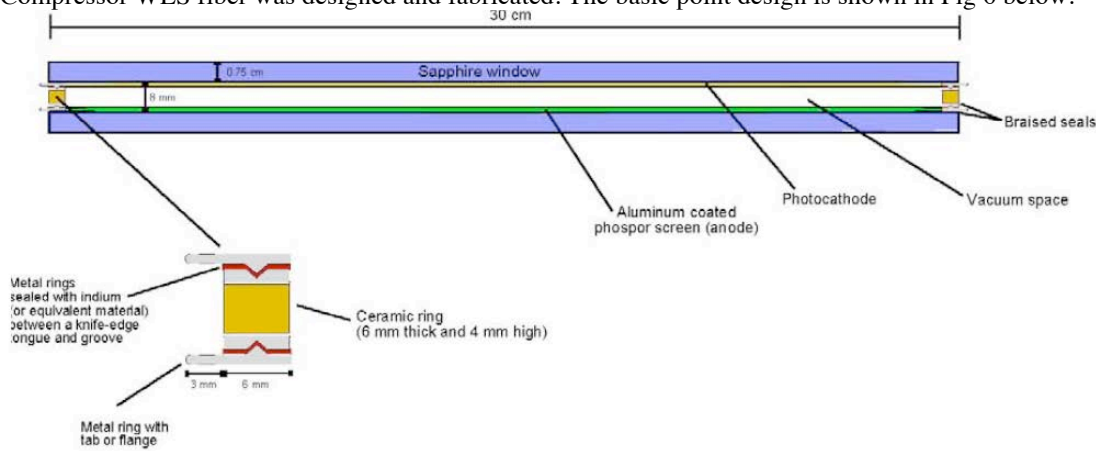

Fig. 6: Design of a large – 30cm diameter – planar (1-D) LAC tube, for matching to a WLS fiber optic output. The design was scaled down to 10 cm (4") in the first realization (described below).

The design of a prototype device using a ceramic annulus for HV and vacuum sealing was scaled from the drawing in Fig. 6 to 4"(10cm). Large planar OCA tubes where the vacuum windows are supported by posts are possible where the loss of signal in the areas taken by the posts are not critical (as might be used in large water Cerenkov underground detectors). The 4" photocathode input and phosphor output windows are synthetic sapphire for strength and UV transmission, including the ZnO:Ga blue/near UV output. A 4" bialkali photocathode deposited on an input window was attached by vacuum transfer to a knife-edge indium seal to a cylindrical planar tube body, with its 4" diameter ZnO:Ga(2%) fast (0.8 ns decay constant) bright phosphor anode, as can be seen in figures 7-11 below. The bi-alkaliphotocathode deposition process was tested first in a separate fabrication test (similar to the description above for the mono-alkali process for the radially focused OCA tube), and in those tests reached an 18% maximum QE on blue light.

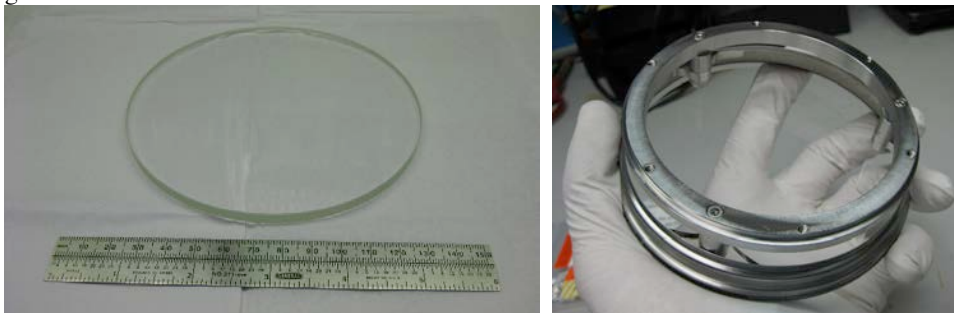

*Fig. 7a: Prefabrication Tests:* The ~4" 7056 alumina input window for photocathode deposition tests, and Phosphor anode window dummy in holder for processing and sealing tests.

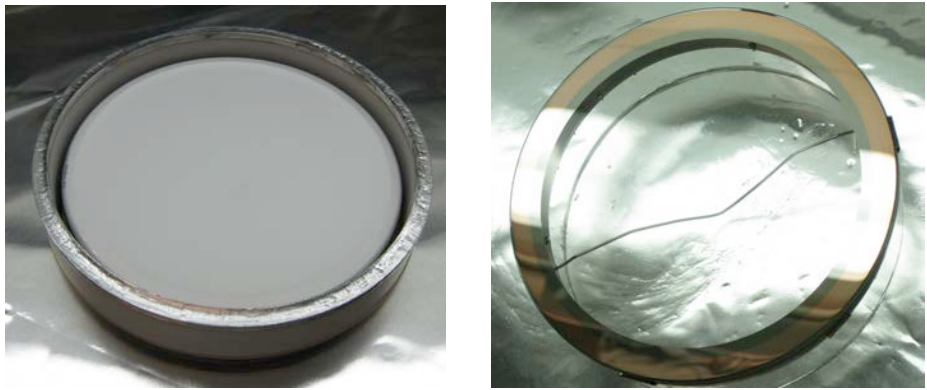

*Fig. 7b: Preforms:* ZnO:Ga Phosphor screen anode(L) and Photocathode window preloads ready to go into the vacuum system(tank) for photocathode formation and sealing to the phosphor screen.

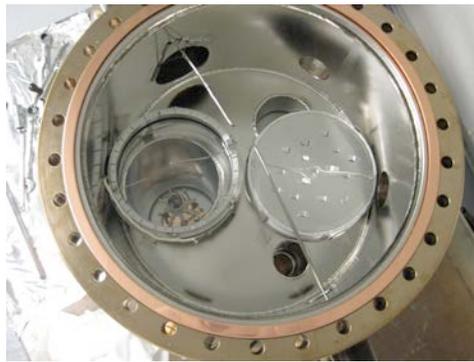

*Figure 8: Assembly/Fabrication Rig:* Vacuum assembly tank for the ZnO:Ga Phosphor screen anode and Photocathode preloads shown above.

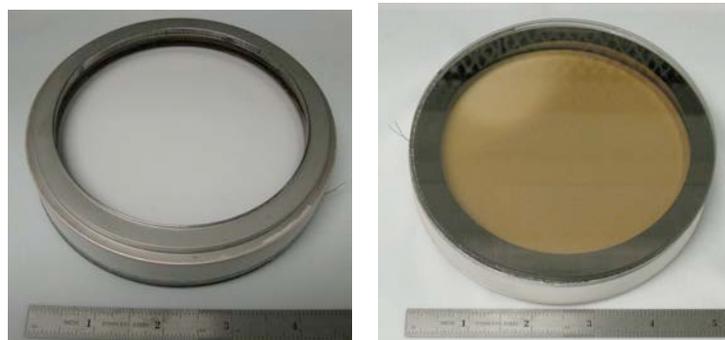

*Figure 9: Finished Planar Optical Compressor Photographs* – Phosphor anode side (Left) and top photocathode(Right).

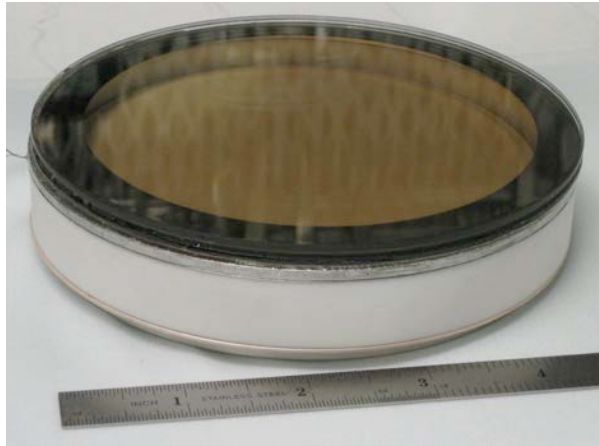

*Fig. 10: Finished Optical Phase Space Optical Compressor/Photon Amplifier:* Side view of finished planar optical phase space compressor/photon amplifier.

The photocathode QE vs wavelength was measured using standard techniques with a monochromator and DC photocurrent and is shown in Figure 11.

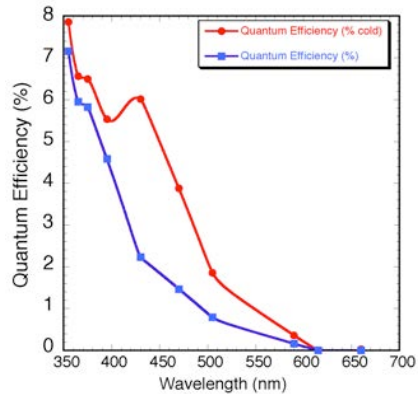

*Figure 11: Photocathode QE:* Measured photocathode QE both at forming temperature (blue) and room temperature(red), using a calibrated monochromator (temperature-compensated commercial Si diode detector) and monitoring the DC photocurrent from the photocathode with a small bias voltage to ensure collection. The uniformity across the 10 cm face is dominated by the cathode uniformity, which is imperfect – about 30% low - over a ~20° "wedge" due to the failure of one of the antimony beads in the evaporation, but the remaining area is uniform to about +/- 8%.

*Fiber Readout:* We constructed 4" diameter 3mm thick UVT lucite transmitting matching plates, optical-grease coupled to the phosphor anode output window. The UV->Blue wavelength shifting (WLS) 1 and 2 mm fibers tested in the grooved matching plates with straight lines, crossed straight lines, and U-shaped grooves. We have simulated the photon propagation from the phosphor screen into the matching plate, with both with specular and diffuse reflectors, and have shown that 7 bounces are the average number to intersect a WLS fiber with a bundle size of ~3mm diameter, an acceptable number of bounces to collect the light using miniature PMT. APD or SiPM may be used instead of PMT to detect the fiber light. A trace of a typical photon in MC is shown below.

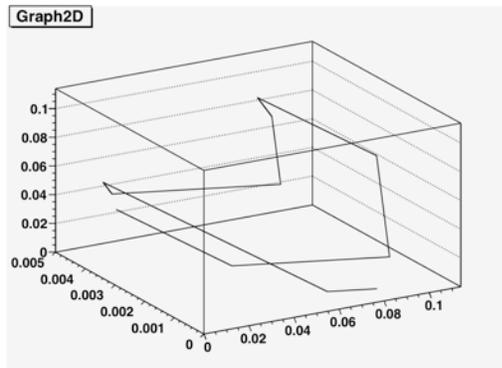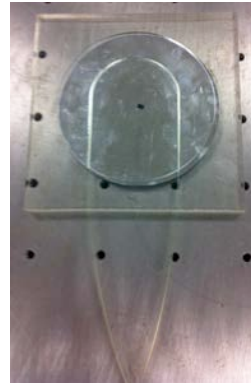

*Fig. 12: Photon Transport MC:* (L) Typical photon tracked from the phosphor output plate to absorption on a WLS fiber. In this case, a single 1 mm fiber was used in a 3mm thick UVT Lucite output matching plate (not to scale) with 9 bounces (95% reflection). Typically the number of bounces scales with the area of fiber in the plate. (R) A fiber readout fiber as MC at left.

A photograph of the test box & apparatus used in the performance scans of the output plate/WLS fibers is shown below in Figure 13.

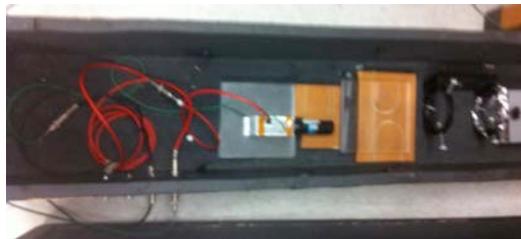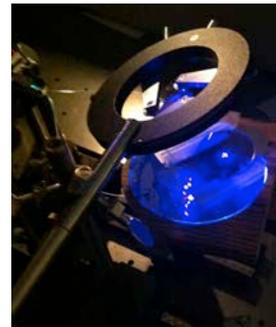

*Fig. 13: WLS Readout Tests:* Test box with UV/Violet laser and PMT (Left) to scan the WLS(wavelength shifting) fiber matching plate attached to the phosphor plate (Right). The response was uniform across the best WLS matching plates to +/-4%. The tests verified efficient light collection from a ZnO:Ga output phosphor plate.
.
*Initial Tests:* The finished Optical Compressor/Amplifier Tube (OCAT) was placed in a dark box and the cathode exposed to the light from a fiber optic cable driven by pulsed LED or a dye laser. Light from the phosphor anode was read-out by 4 x 1 mm diameter green WLS fibers as shown figure 14. This planar optical compressor/photon amplifier tube was operated with the photocathode at ground, and the anode at +HV. The HV was between 10KV and 60KV. The response (arbitrary units) was linear within 7% from 12KV - 40 KV, using a 425 nm LED.

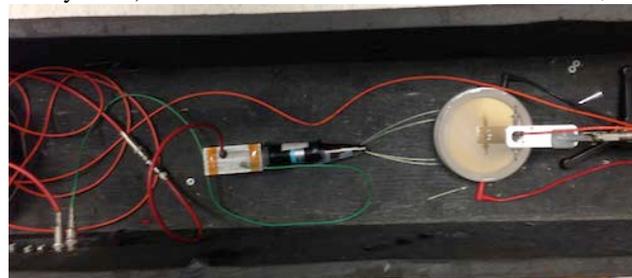

*Fig. 14: Dark Box Test Fixture:* On the far right is an optical fiber in a mounting fixture, pointed down on the photocathode of the optical compressor amplifier. The red quartz fiber with SMA connectors feeding through the dark box from an external laser or LED is shown across the top of the photograph. Four WLS(wavelength shifting) fibers can be seen reading out the phosphor anode, just to the right of the middle. A small PMT is reading out the 4 WLS fibers just at the left of the middle. The 40 KV HV power is shown with the red plug connector at bottom.

The pulse risetime was 4.5 ns, a reasonable value for a 4" tube less than 1" thick, readout by WLS fibers, with a green or blue LED pulse width after fiber transmission ~12 ns, and a small PMT with a 2 ns risetime. We estimate the photon gain at the end of the readout fibers is between 2-4 photons out per incident photon in at the wavelength tested and at 40KV (425 nm is fairly optimal for water Cerenkov transmission over 10 meters in water). This is below expected performance; the low value of the QE compared to standard bialkali is the reason. If the performance improves to 6 photons/photon gain, then with modern small high QE PMT or SiPM, we can expect 2-3 p.e. per incident photon.

The photon phase space areal compression of (4" diameter photocathode area)/ (4 x 1mm diameter fibers) ~ 8,000. The angular phase space compression is estimated to be a factor of 4 using the NA of the fiber. Thus the planar OCA tube offers an interesting option for large area photodetectors for large water calorimeters – which could use posts to support large planar photocathodes in devices as large as 30-100 cm major diameters.

*Conclusion:*
We have demonstrated a simple idea for detecting light over large areas, based on vacuum photocathodes, vacuum photodiodes and Gen-0 image intensifiers, by making a "light-amplifier" and collecting the light on optical fibers. Phase space compression factors angle x area exceeding 10,000 have been demonstrated, with amplifications of 2-4 photons per incident photon, and pointing to the possibility of considerably more. The time resolution possible with a 4-5 ns risetime pulse is reasonable, as is the potential repetition rate, as the phosphor recovers in sub-ns times.

*Major Advantages of an Optical Amplifier/Compressor*
We would anticipate manufacturable fiber-optic output photoamplifiers. The potential advantages of devices of this kind are:

a) no (or very low power) base HV divider string (i.e. a base only for focussing electrodes if necessary);
b) no cable driving amplifiers necessary;
c) very low power at HV;
d) can couple many large-area photodetectors to a single multi-pixel detector;
e) can be highly compact; large photocathode area per volume of vacuum tube;
f) simple construction - one scintillating fiber or plate & no dynodes - (we would anticipate a low cost per unit);
g) good radially focused, proximity focused, or demagnified electron optics and collection-with a sub-ns phosphor should have good jitter characteristics;
h) small cable cross-section;
i) noise immunity on the fiber-optic output from power ripple and external electromagnetics;
j) very good photon gain stability and tube-tube gain uniformity for ease of calibration;
k) excellent radiation hardness, especially with quartz-only fiber cables and envelopes;
l) excellent optical pulse linearity;
m) gain linear with voltage, as contrasted with a photomultiplier, for modest voltage stability requirements (i.e. ripple can be 0.5% and maintain 0.5% gain stability);
n) Operation in multi-Tesla axial magnetic fields, in some tube configurations.
o) Very low radiation-induced backgrounds
p) Scales to very large sizes in the cylindrical form factor (like fluorescent light tubes) or in the planar factor (support posts between the photocathode window and the phosphor anode are ok in this technology), with good capability of pressure standoff especially in the cylindrical configuration for underwater or large underground water Cerenkov detectors.

*Acknowledgements*: We thank the NSF for support under Grant DUSEL Award Number NSF 081064, and the U.S. DOE for subsidiary funding through US-CMS. We thank the Physics Department of Fairfield University for support and technical help, Ossy Siegmund for design and fabrication help in all phases of the planar OCA tube, and High Energy Physics Lab at Harvard University for help with the cylindrical tube (W.A.Huffman and A. Sommer).

---